\documentclass[twocolumn,showpacs,preprintnumbers,amsmath,amssymb,
showkeys]{revtex4}

\usepackage{graphicx}
\usepackage{dcolumn}
\usepackage{bm}
\usepackage{bigints}

\begin{document}

\title{Anisotropic thermal Sunyaev-Zel'dovich effect and the
  possibility of an independent measurement of the CMB dipole, quadrupole and octupole}

\author{I.G. Edigaryev}
\affiliation{Astro-Space Center of P.N. Lebedev Physical Institute, Profsoyusnaya 84/32, Moscow, Russia 117997.}

\author{D.I. Novikov}
\affiliation{Astro-Space Center of P.N. Lebedev Physical Institute, Profsoyusnaya 84/32, Moscow, Russia 117997.}

\author{S.V. Pilipenko}
\affiliation{Astro-Space Center of P.N. Lebedev Physical Institute, Profsoyusnaya 84/32, Moscow, Russia 117997.}


\begin{abstract}
  We consider the effect of the cosmic microwave background (CMB) frequency spectral distortions
  arising due to the Compton scattering of the anisotropic radiation on
  Sunyaev-Zel'dovich (SZ) clusters. We derive the correction to the thermal SZ 
  effect due to the presence of multipoles with $\ell=1,2,3$ in the
  anisotropy of the CMB radiation. 
  We show that this effect gives us an opportunity for an independent
  evaluation of the CMB dipole, quadrupole and octupole angular anisotropy
  in our location using distorted signal from the nearby galaxy clusters and
  to distinguish between the Sachs–Wolfe (SW) and the Integrated Sachs–Wolfe (ISW)
  effects by combining such signals from distant and nearby clusters.
  The future space mission 'Millimetron' will have unprecedented sensitivity,
  which will make it possible to observe the spectral distortion we are
  considering.
\end{abstract}

\keywords{Cosmic Microwave Background, spectral distortions, SZ effect,
 cosmology}

\maketitle

\section{Introduction}

In this paper we propose a method for the independent estimation of the Cosmic Microwave Background (CMB) dipole, quadrupole and octupole anisotropy using measurements of the relic radiation frequency spectrum distortion from nearby and distant Sunyaev-Zel'dovich clusters.

In the past two decades, a large number of papers, devoted to the theoretical study of various corrections to the Kompaneets equation \citep{1957JETP...4.730K} and the Sunyaev-Zel'dovich effect \citep{1969Ap&SS...4..301Z,1970Ap&SS...7...20S} have been published. These studies take into account relativistic corrections to the thermal SZ effect \citep{1998ApJ...499....1C,1998ApJ...502....7I,Stebbins:1997qr,Rephaeli:2002zs} and multiple scattering
on SZ
clusters \citep{2000astro.ph..5390I}. Recently in Ref.~\citep{2018arXiv180303277H} the temperatures of galaxy clusters were measured for the first time using the relativistic corrections to the thermal SZ effect only. The relativistic correction approach was adopted in Refs.
\citep{1998ApJ...508...17N,1998ApJ...508....1S,1999ApJ...510..930C} for moving clusters of galaxies to find corresponding corrections for the kinematic SZ effect. In
   Ref. \citep{2000ApJ...533..588I} the covariant formalism
for the polarization SZ effect  was applied. In Refs. \citep{2012MNRAS.426..510C} a way to cleanly separate kinematic
and scattering terms was obtained. In Ref. \citep{2009PhRvD..79h3005N,2013MNRAS.434..710N,2014MNRAS.441.3018N}, a detailed analytic investigation
of the Boltzmann equations was made for three Lorentz frames and expressions for the photon redistribution functions were derived. As for the study of the intensity distortions, the basic assumption of these works is the CMB is isotropic.

The influence of the CMB anisotropy on the spectral distortions was investigated in Refs. \citep{2016PhRvD..94b3513Y,Balashev:2015lla} for the kinetic Sunyaev-Zel'dovich effect. In that work polarization
and intensity distortions in moving clusters were considered, but for simplicity the thermal effect was not taken into account.
In Ref. \citep{2017PhRvL.119v1102Y} it was demonstrated how
spectral measurements of the low multipoles can be used to separate the motion-induced dipole
of the CMB from a possible intrinsic dipole component.

The generalized Kompaneets equation was considered in Ref.
\citep{2012MNRAS.425.1129C,2014MNRAS.437...67C,2014MNRAS.438.1324C}.

In our paper we consider the intensity spectral distortion due to Compton scattering of anisotropic CMB radiation on nonmoving galaxy clusters. We do not take into account the movement of clusters since the effect under consideration can be easily separated from the effects associated with clusters' peculiar
velocities.

Compton scattering of an anisotropic blackbody radiation on plasma causes a very characteristic spectral distortion due to the presence of  multipoles with $\ell=1,2,3$ in the radiation anisotropy. This effect is the anisotropic correction to the classical thermal SZ effect and follows directly from the anisotropic Kompaneets equation first derived in Ref. \citep{1969JPhys..30..301B} in linear approximation in $kT_e/m_ec^2$. The amplitude of this distortion depends on the powers of $C_1, C_2, C_3$ multipoles and their orientation with respect to the axis connecting the scattering point and the observer. We will show, that the shape of this distortion is very characteristic and allows to distinguish it from other nonblackbody components.

The CMB radiation for the observer located at a nearby SZ cluster has roughly the same anisotropy map as the radiation that we directly observe in the sky including the quadrupole and octupole. We will show, that for redshifts
$z\sim 0.05$ the variations of multipole
amplitudes are about $10\%$. Therefore, the measurements of the spectral distortions from such clusters caused by the presence of dipole, quadrupole and octupole anisotropy at the moment of scattering gives us an opportunity
for an independent estimation of the $\ell=1,2,3$ CMB multipoles and
their orientations.  According to WMAP and Planck results, the low CMB anisotropy multipoles have insufficient power (compare to the expected one) \citep{2003MNRAS.346L..26E,2003PhRvD..68l3523T,2004PhRvL..93v1301S,2014A&A...571A..15P}. Besides there is a quadrupole-octupole alignment \citep{2004PhRvD..70d3515C,2006MNRAS.367...79C,2008IJMPD..17..179N}. These facts are rather difficult to explain in the framework of Gaussian statistics. As for the CMB intrinsic dipole it is
completely overshadowed by our motion with respect to the CMB frame.
Thus, an independent estimate of the powers of quadrupole and octupole (our local multipoles we directly observe by WMAP and Planck) and their orientations can be quite important. Besides it can provide us with the information about the intrinsic dipole in our location.

At the same time, the use of such a signal from distant clusters can provide us with unique
information about the pure Sachs–Wolfe (SW) effect without the integrated part of it which
is the Rees-Sciama (RS) \citep{1967ApJ...147...73S} or Integrated Sachs-Wolfe (ISW) effect. The ISW effect
creates its contribution to the CMB low multipoles at very small redshifts. Therefore the
probe of the spectral distortions caused by the anisotropic SZ effect from high redshift
SZ clusters is ISW-free.

The upcoming Millimetron mission \citep{2009ExA....23..221W,2014PhyU...57.1199K,
  2012SPIE.8442E..4CS} will have unprecedented sensitivity ($\sim$ 1 Jy/Sr) in the single dish mode to measure the signal in the range from 60 GHz to 1 THz. One of the main tasks of this mission will be the measurement of so called $\mu$- and $y-$ spectral distortions of CMB radiation, which arise due to the energy injections into the plasma in the prerecombination epoch (see for instance Refs.
\citep{2015MNRAS.451.4460D,2014arXiv1405.6938C}). As an additional task, the possibility of detecting the spectral distortions from SZ clusters caused by the presence of CMB anisotropy can be considered.

As it was mentioned above, in the approximation in $kT_e/m_ec^2$
only the first three multipoles of the CMB anisotropy (dipole, quadrupole and octupole) form the additional spectral distortion in the thermal SZ effect. In order to separate the signal, caused by the presence of these
multipoles at the moment of scattering from other spectral distortions it is necessary to take into account all the effects of higher or the same
order for the isotropic (monopole) CMB fraction. These effects include relativistic corrections to the thermal SZ effect at least up to the fifth order
in $kT_e/mc^2$, corrections to the kinetic SZ effect and the effect of multiple scattering.

The paper is organized as follows: In Section II we review the equation for the radiative transfer in plasma for relatively cold electrons and photons for the
linear order in $h\nu/m_ec^2$, $kT_e/m_ec^2$. Using this equation we derive the
correction to the thermal SZ effect due to the presence of dipole,
quadrupole and octupole anisotropies in CMB radiation. In Section III we
 present the method for independently estimating the $\ell=1,2,3$ multipoles
 in our location using nearby SZ clusters. In this Section we also
 show how to distinguish between SW and ISW effects combining the signals from
 nearby and distant clusters. Finally in Section IV we make our conclusions.
 
\section{The anisotropic Kompaneets equation and the anisotropic thermal SZ effect}
In this Section we review the anisotropic Kompaneets equation found by
Babuel-Peyrissac and G. Rouvillois \citep{1969JPhys..30..301B} and derive the formula for the
anisotropic thermal SZ effect.

We use the following notations:

\vspace{0.3cm}

$I=I(\nu)$: spectral radiance;

\vspace{0.3cm}
$n(\nu,\bold{\Omega})=\frac{c^2I}{2h\nu^3}$: photon concentration in phase
space (photon number density), where $\nu, c, h$  are the frequency, the speed of light and the Planck constant respectively;

\vspace{0.3cm}
$\bold{\Omega'}$, $\bold{\Omega}$: radiation propagation directions
before and after scattering;

\vspace{0.3cm}
$\mu=\bold{\Omega}\bold{\Omega'}$: cosine of the scattering angle;

\vspace{0.3cm}
$T_e, T_r$: temperatures of electrons and radiation;

\vspace{0.3cm}
$m_e$: electron mass at rest;

\vspace{0.3cm}
$N_e$: concentration of electrons;

\vspace{0.3cm}
$\Theta_e=\frac{kT_e}{m_ec^2}$, where $k$ is the Boltzmann constant;

\vspace{0.3cm}
$\sigma_{T}$: Thomson cross section;

\vspace{0.3cm}
we also use the operator $\frac{D}{D\tau}=\frac{1}{N_e\sigma_T}\frac{d}{cdt}+\bold{\Omega}\nabla$ and the variable $x=\frac{h\nu}{kT_r}$.

\vspace{0.3cm}
The equation for the radiative transfer in plasma was derived in Ref. \citep{1969JPhys..30..301B}
for relatively cold photons and electrons: $kT_e\ll m_ec^2$,
$h\nu\ll m_ec^2$. This approximation is quite correct for the typical
SZ clusters with $T_e\sim 10\hspace{0.3cm}KeV$ and relic photons with $T_r\sim 3^{\circ}K$. The equation we consider, describes the correction to the Thomson scattering up to the first order in $h\nu/m_ec^2$, $kT_e/m_ec^2$.

In order to separate the anisotropic and isotropic parts of radiation
we denote by $\Delta$ the difference between the photon number density before and after scattering:
$\Delta=n(x,\bold{\Omega'})-n(x,\bold{\Omega})$. Following Ref. \citep{1969JPhys..30..301B} and using
our notations one can rewrite the equation for the radiative transfer in
plasma in terms of $n(x,\bold{\Omega})$:
\vspace{-0.1cm}
\begin{equation}
  \begin{array}{l}
    
    \vspace{0.5cm}
    
    \frac{Dn}{D\tau}=
    \frac{3}{16\pi}\bigint\limits_{\bold{\Omega'}}
    \Big\{\left.(1+\mu^2)\Delta+\right.\\

     \vspace{0.5cm}

     \left.
+\Theta_e\left(4\mu^3-6\mu^2-4\mu+2-2\frac{T_r}{T_e}x(1+\mu^2)(1-\mu)
  \right)\right.\Delta+\\
  
  \vspace{0.5cm}

  \left.
 + \Theta_e\left(\frac{1}{x^2}\frac{\partial}{\partial x}
  \left[ x^4\left(\frac{T_r}{T_e}\Delta+
    \frac{\partial\Delta}{\partial x}\right)\right]
  +2\frac{T_r}{T_e}n
  \frac{\partial}{\partial x}\left[x^2\Delta\right]\right)\cdot\right.\\

   \vspace{0.5cm}

  \left.
  \cdot(1+\mu^2)(1-\mu)\right.\Big\}
  d\bold{\Omega'}+\\

\vspace{0.5cm}

+\Theta_e\frac{1}{x^2}\frac{\partial}{\partial x}\left[x^4\left(\frac{T_r}{T_e}
  (n+n^2)+\frac{\partial n}{\partial x}\right)\right]+O(\Theta_e^2),
 \end{array}
\end{equation}
where $n=n(x,\bold{\Omega})$.
The integral part of this equation represents the contribution from the anisotropic radiation, while the last term gives the usual right hand side of the
Kompaneets equation. Taking into account that $T_r/T_e$ is extremely
small for SZ clusters, we neglect all terms proportional to this ratio:

\begin{figure}[tbh]
  \includegraphics[width=1\columnwidth]{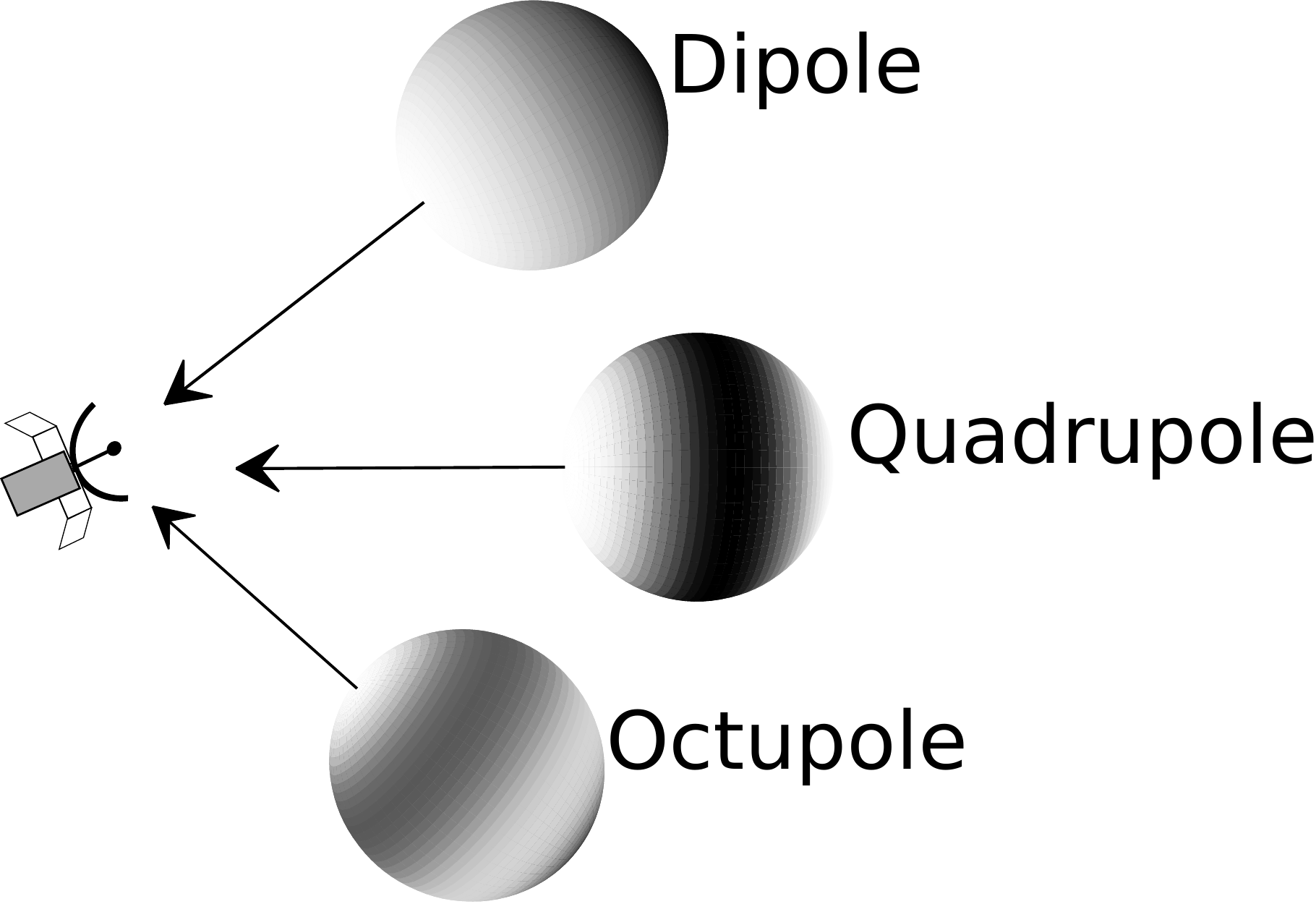}
  \caption{Schematic representation of multipole components which affect the observed CMB spectrum due to scattering. Only components with
$\ell=1,2,3$ and with the symmetry axis directed towards an observer contribute to the spectral distortion.}
\end{figure}

\begin{equation}
  \begin{array}{l}
    
    \vspace{0.5cm}
    
    \frac{Dn}{D\tau}=\frac{3}{16\pi}\bigint\limits_{\bold{\Omega'}}
    (1+\mu^2)\Delta d\bold{\Omega'}+\\

    \vspace{0.5cm}
    
    +\Theta_e\frac{3}{16\pi}\bigint\limits_{\bold{\Omega'}}\Big\{
    \left(4\mu^3-6\mu^2-4\mu+2\right)\Delta+\\
    
    \frac{1}{x^2}\frac{\partial}{\partial x}
  \left[ x^4\frac{\partial}{\partial x}(n+\Delta)\right]
  (1+\mu^2)(1-\mu)\Big\}d\bold{\Omega'}

 \end{array}
\end{equation}

The cosmic microwave background radiation has a spectrum that coincides with great accuracy with the blackbody spectrum, defined by the temperature $T_r$. Since the relic radiation is anisotropic, its temperature varies with direction. Thus, for radiation incident from the direction
$\bold{\Omega'}$, we can write the following formula:

\begin{equation}
  \begin{array}{l}
    
    \vspace{0.5cm}
    
    n(x,\bold{\Omega'})=B+\frac{dB}{dx}\hspace{0.1cm}\frac{dx}{dT_r}
    \hspace{0.1cm}\Delta_T(\bold{\Omega'}),\\

    \vspace{0.5cm}
    
    B(x)=\frac{1}{e^x-1},
    \end{array}
\end{equation}
where $\Delta_T(\bold{\Omega'})=T(\bold{\Omega'})-T_r$.
Therefore, in order to find the spectral distortions that arise during a single scattering of photons on plasma in SZ clusters, we substitute the following initial conditions into Eq. (2):

\begin{figure}[tbh]
  \includegraphics[width=1\columnwidth]{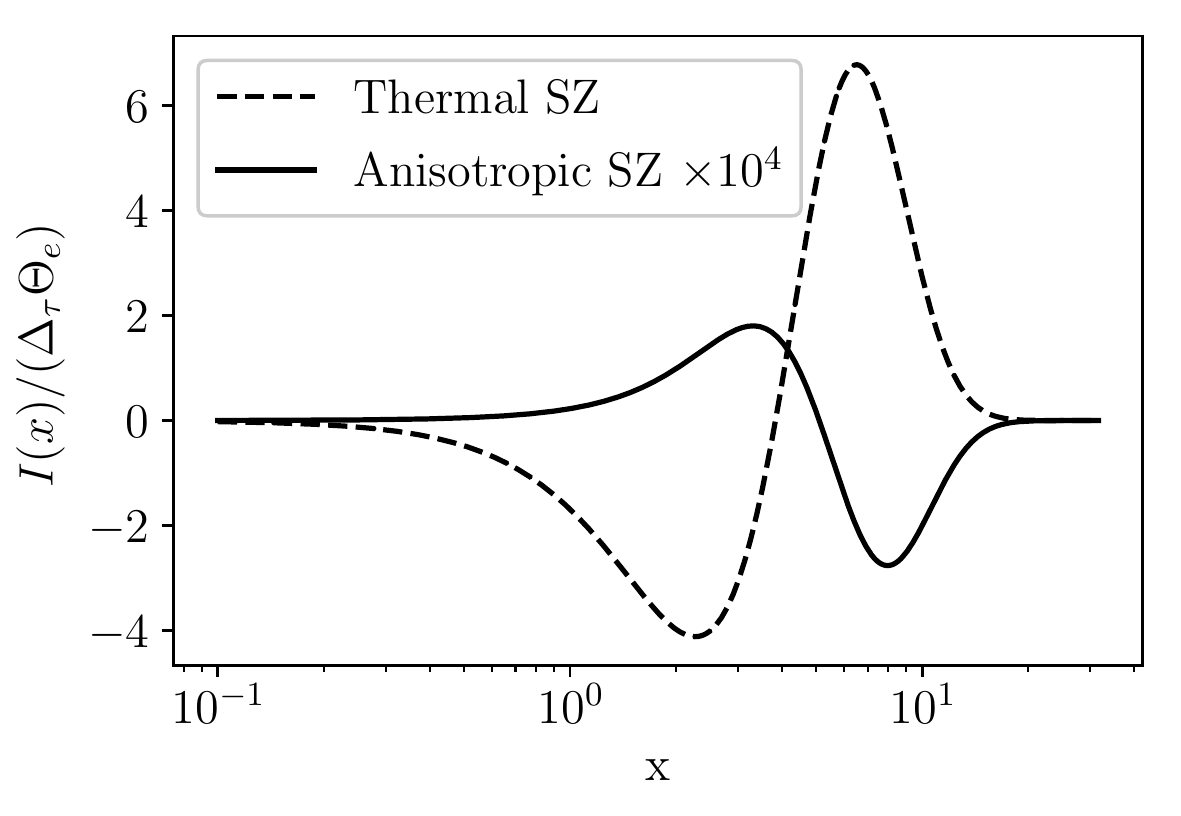}
  \caption{The comparison of the classical thermal SZ effect (dashed line)
    with the spectral distortion caused by the presence of $\ell=1,2,3$
    anisotropy multipoles (Anisotropic thermal SZ effect) (solid line).}
\end{figure}

\begin{equation}
  \begin{array}{l}

    \vspace{0.5cm}
    
  n(x,\bold{\Omega'})=n+\Delta=B-x\hspace{0.1cm}\frac{dB}{dx}\hspace{0.1cm}
  \frac{\Delta_T(\bold{\Omega'})}{T_r},\\

  \vspace{0.5cm}
  
  \Delta=-x\hspace{0.1cm}\frac{dB}{dx}\hspace{0.1cm}
  \left(\frac{\Delta_T(\bold{\Omega'})}{T_r}-
  \frac{\Delta_T(\bold{\Omega})}{T_r}\right)
  \end{array}
\end{equation}
Finally we get the result for the changes in photon number
density $\Delta_n$ in the optically thin limit:
\begin{equation}
  \begin{array}{l}

    \vspace{0.5cm}
    
    \Delta_n/\Delta_{\tau}=\\

    \vspace{0.5cm}
    
    =(\beta_1+\Theta_e\beta_2)
    \hspace{0.1cm}\left(-x
    \hspace{0.1cm}\frac{dB}{dx}\right)+\hspace{1.7cm}\Big\}1\\

    \vspace{0.5cm}
        
    +\Theta_e\beta_3\hspace{0.1cm}\frac{1}{x^2}\hspace{0.1cm}
    \frac{d}{dx}\left[x^4\frac{d}{dx}
      \left(-x\frac{dB}{dx}\right)\right]+\hspace{1cm}\Big\}2\\

    \vspace{0.5cm}
    
    +\Theta_e\hspace{0.1cm}\frac{1}{x^2}\hspace{0.1cm}\frac{d}{dx}
    \left[x^4\frac{dB}{dx}\right],\hspace{2cm}\Big\}tSZ
    
  \end{array}
\end{equation}
where $\Delta_{\tau}$ is the optical depth. The first two terms
in Eq. (5) are due to the presence of non-zero
CMB anisotropy. The physical meaning of these terms is given below.
The third term represents the classical thermal SZ effect for the isotropic
fraction of radiation.
According to
Eq. (2) expressions for the coefficients $\beta_{i}$ are as follows:
\begin{equation}
  \begin{array}{l}

    \vspace{0.5cm}
    
   \beta_1=\frac{3}{16\pi}\bigint\limits_{\bold{\Omega'}}
    (1+\mu^2)\left(\frac{\Delta_T(\bold{\Omega'})}{T_r}-
   \frac{\Delta_T(\bold{\Omega})}{T_r}\right) d\bold{\Omega'},\\

   \vspace{0.5cm}
   
   \beta_2=\frac{3}{16\pi}\bigint\limits_{\bold{\Omega'}}
    (4\mu^3-6\mu^2-4\mu+2)\left(\frac{\Delta_T(\bold{\Omega'})}{T_r}-
   \frac{\Delta_T(\bold{\Omega})}{T_r}\right) d\bold{\Omega'},\\

   \vspace{0.5cm}
   
   \beta_3=\frac{3}{16\pi}\bigint\limits_{\bold{\Omega'}}
   (1-\mu)(1+\mu^2)\frac{\Delta_T(\bold{\Omega'})}{T_r}
   d\bold{\Omega'}.
    
  \end{array}
\end{equation}
For the observer located at the SZ cluster the CMB radiation anisotropy
can be described in terms of spherical harmonics $Y_{\ell}^m(\bold{\Omega'})$:

\begin{equation}
  \frac{\Delta_T(\bold{\Omega'})}{T_r}=\sum\limits_{\ell=1}^{\infty}
  \sum\limits_{m=-\ell}^{\ell}\tilde{a}_{\ell m}Y_{\ell}^m(\bold{\Omega'}),
\end{equation}
where $\tilde{a}_{\ell m}$ are the coefficients for $\Delta T/T$ decomposition
for a given location of the galaxy cluster.
If we choose a local spherical coordinate system at the scattering point in such a way that the north pole points to the observer [which is the same as the direction $\bold{\Omega}$ in Eqs. (1) and (2)], then in terms of such a system the coefficients $\beta_i$ are as follows:

{\large
\begin{equation}
  \begin{array}{l}

    \vspace{0.3cm}
    
    \beta_1=\frac{1}{4\sqrt{\pi}}\left(\frac{1}{\sqrt{5}}\tilde{a}_{20}
     -2\frac{\Delta_T(\bold{\Omega})}{T_r}\right),\\

    \vspace{0.3cm}
    
    \beta_2=-\frac{3}{2\sqrt{\pi}}\left(
    \frac{2}{5\sqrt{3}}\tilde{a}_{10}+\frac{1}{\sqrt{5}}\tilde{a}_{20}
    -\frac{2}{5\sqrt{7}}\tilde{a}_{30}\right),\\

    \vspace{0.3cm}
    
   \beta_3=-\frac{3}{4\sqrt{\pi}}\left(\frac{4}{5\sqrt{3}}\tilde{a}_{10}-
    \frac{1}{3\sqrt{5}}\tilde{a}_{20}+
    \frac{1}{5\sqrt{7}}\tilde{a}_{30}\right).

  \end{array}
\end{equation}
\par}
We use tilde for $\tilde{a}_{\ell m}$ in Eqs. (7) and (8) because these
coefficients are not the same as the conventional $a_{\ell m}$ related to our location
and galactic coordinate system. In Fig 1 you can find a schematic representation of multipole components which contribute to the observed CMB frequency spectrum distortions due to scattering.

One can notice from Eqs. (5) and (8)
the following features of the formation of spectral distortions due to Compton
scattering of anisotropic radiation. 

\vspace{0.3cm}
1. The distortion of the form $-x\frac{dB}{dx}$ [term 1 in Eq. (5)] arises 
as a result of mixing flows with different temperatures due to Thomson
scattering. Since $\Theta_e\ll 1$, the amplitude of this distortion
is dominated by $\beta_1$. This coefficient consists of two terms.
The first one is the component $\tilde{a}_{20}$ of the
local quadrupole temperature anisotropy. It is worth noticing, that this
is the only quadrupole component, which does not produce polarization.
The second term has the following physical meaning. If there is a hot spot in the
CMB anisotropy behind the SZ cluster, then hotter photons, propagating towards the observer through the SZ cluster, are partially scattered out of the
line of sight and replaced by cooler photons scattered in the line of sight. As
a result of this effect, we would observe a blackbody radiation with a bit
lower temperature than in the absence of SZ cluster. Therefore, both effects
change the temperature of radiation. Indeed, the term
$\frac{\Delta_T(\bold{\Omega})}{T_r}\hspace{0.1cm}x\frac{dB}{dx}$
is in fact the
difference between two blackbodies' spectra,
$B(T_r)-B(T_r+\Delta_T(\bold{\Omega}))$,
in linear approximation. That's why it is rather
difficult to use distortion
of this kind to find the $\tilde{a}_{20}$: this effect simply changes the temperature, leaving the spectrum in a blackbody shape.
\vspace{0.3cm}

2. The deviation of the form $\frac{1}{x^2}\hspace{0.1cm}
    \frac{d}{dx}\left[x^4\frac{d}{dx}
      \left(-x\frac{dB}{dx}\right)\right]$ [term 2 in Eq. (5)]
    has a very distinctive
    nonblackbody features. This is the deviation, that we call the
    ``Anisotropic thermal SZ effect''. The amplitude of this deviation is
    proportional to the linear
    combination of the local dipole, quadrupole and octupole components only.
    In Fig. 2 we demonstrate this distortion in comparison with the usual
    thermal
    SZ effect. In the next section we show how to use these spectral features
    in the radiation coming from nearby SZ clusters to independently
    estimate the $a_{\ell m}$, ($1\le\ell\le3$, $-\ell\le m\le\ell$)
    components, that we directly observe on the sky. We also
    show how to distinguish between the SW and ISW effects combining the signals from
    nearby and distant clusters.

    \section{The estimation of the CMB ${\ell=1,2,3}$ anisotropy multipoles
      and separation of SW and ISW effects}

    In this Section we describe the method to reconstruct the CMB dipole, quadrupole
    and octupole at our location and show how to separate contributions from the
    SW and ISW effects to CMB anisotropy by observing
    the distorted signals from SZ galaxy clusters. The outline of this section is as follows.
    In Sec. III A we demonstrate the possibility of $a_{\ell m}$ reconstruction for
    the simplest possible model and an ideal experiment with no noise. In Sec. III B we
    consider the $a_{\ell m}$ spatial autocorrelations caused by SW and ISW effects and propose
    a method for separating SW and ISW contributions to the total $a_{\ell m}$. Finally in
    Sec. III C we make our estimates
    of relative errors for different $\ell$ and $m$ using SZ clusters from the Planck catalog.
    
 \subsection{The possibility of $a_{lm}$ reconstruction}

    We start our analysis with a very simple and visual approximation. Let us assume
    for simplicity, that the CMB incident to all observed SZ clusters has exactly
    the same anisotropy.

Let us denote by $a_{\ell m}^j$ coefficients of the temperature anisotropy
decomposition for the observer located at cluster number $j$.
Taking into account our simple approximation we consider these coefficients as equal to those
we directly observe in the sky: $a_{\ell m}^j=a_{\ell m}$.
This gives us an opportunity to estimate $a_{\ell m}$, $\ell=1,2,3$,
$-\ell\le m\le \ell$ by observing the CMB radiation spectral distortions coming
from such clusters. The relations between $a_{\ell m}$ of the conventional
galactic coordinates and
$\tilde{a}_{\ell m}$ defined in Eq. (7) are given by the rotation of the coordinate system. This rotation transforms a spherical harmonic of degree $\ell$ and
order $m$ into a linear combination of spherical harmonics of the same degree.
Therefore the $\tilde{a}_{\ell 0}^j$ coefficients can be expressed in terms
of $a_{\ell m}$ as follows:

\begin{equation}
  \tilde{a}_{\ell 0}^j=\sum\limits_{m=-\ell}^\ell D_{\ell, j}^{0,m}a_{\ell m}^j
  \approx\sum\limits_{m=-\ell}^\ell D_{\ell, j}^{0,m}a_{\ell m} , 
\end{equation}
where $D_{\ell,j}^{m,m'}$ are the complex conjugate of the Wigner
D-matrix elements for the $j$-th cluster. These elements are completely
described by the position on the sky of the cluster number $j$.
According to Eqs. (5) and (8) the 
\vspace{0.2cm}
ratio of the amplitude of
$\frac{1}{x^2}\hspace{0.1cm}\frac{d}{dx}\left[x^4\frac{d}{dx}\left
  (x\frac{dB}{dx}\right)\right]$
signal to the classical SZ thermal effect is

\begin{equation}
  \begin{array}{l}
    
  \vspace{0.3cm}
  
  s_j=-\frac{3}{5\sqrt{3\pi}}\sum\limits_{m=-1}^1D_{1,j}^{0,m}a_{1m}+
  \frac{1}{4\sqrt{5\pi}}\sum\limits_{m=-2}^2D_{2,j}^{0,m}a_{2m}-\\

  \vspace{0.3cm}
  
  -\frac{3}{20\sqrt{7\pi}}\sum\limits_{m=-3}^3D_{3,j}^{0,m}a_{3m},
  \hspace{0.8cm}j=1,..,N.
  \end{array}
\end{equation}
In total 15 coefficients $a_{\ell m}$ with $\ell=1,2,3$ and
$-\ell\le m\le \ell$
contribute to $s_j$. This means, that at least 15 SZ objects should
be used to solve the linear system (10). Despite the fact that this is a
mathematical certainty, in reality the use of 15 clusters only can lead
to a completely wrong result because the contribution to the signal from the
octupole is overshadowed by the dipole and quadrupole [see Eq. (8), coefficient $\beta_3$].
This means, that even a small noise can lead to an incorrect reconstruction of $a_{3m}$.
Besides, as we show in the next subsection, there are spatial fluctuations of
$a_{\ell m}$ mainly due to the ISW effect. Therefore, even in the case of an ideal experiment
we need many more clusters to perform the reconstruction.
In Fig. 3 we show the expected signal $s_j$ from nearby clusters without the dipole
contribution in case where our measurements of the quadrupole and octupole by
COBE, WMAP and Planck contain information about the real cosmological signal.

\begin{figure}[tbh]
  \includegraphics[width=1\columnwidth]{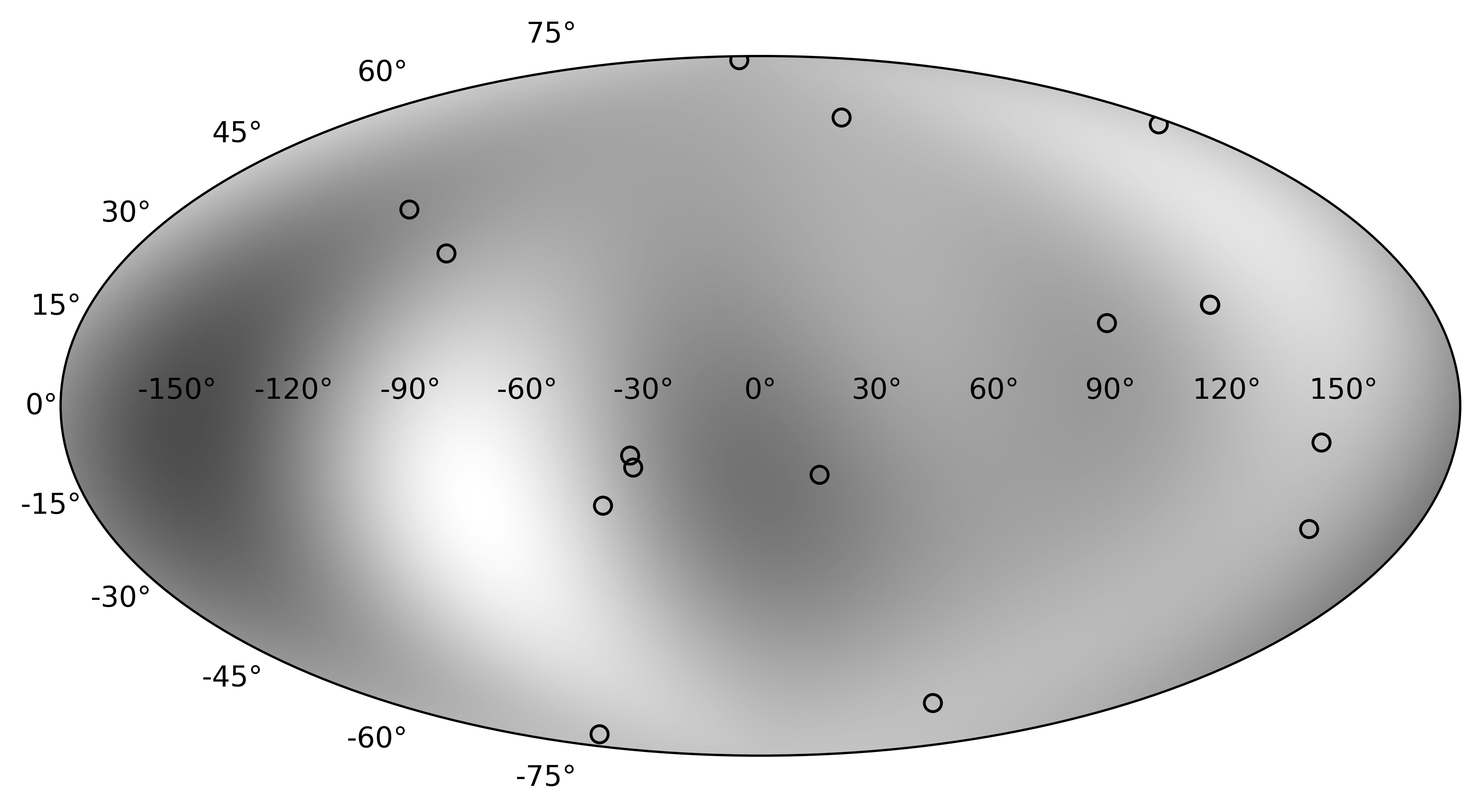}
  \caption{Map of the expected distribution of the distorted signal on the sky (galactic coordinates) coming from nearby SZ clusters. This map is constructed from the quadrupole and octupole measured by Planck. The unknown dipole is not taken into account. Circles represent 15 nearest clusters from PLANCKSZ2 catalog.}
\end{figure}

\subsection{ISW and SW effects and ${a_{\ell m}}$ spatial correlations}

According to Ref. \cite{1996PhRvL..76..575C}, for $\Omega_\Lambda=0.7$ on average 40\% of the quadrupole amplitude is generated by the ISW effect, while for the octupole this quantity reaches 25\%. The rest is caused by the SW effect on the last scattering surface. In order to estimate the spatial variations of both effects, we simulate a Gaussian distribution of a large scale gravitational potential in a 40~Gpc/h cubic box with $512^3$ grid cells in accordance with the known cosmological power spectrum. To compute the ISW effect, we convolve this field with the kernel taken from Ref. \cite{1996PhRvL..76..575C} for $\ell=1,2,3$ and $m=0$. Thus we obtain maps of $a_{\ell 0}$ generated by the ISW effect at the same local time everywhere. From these maps we compute the correlation of $a_{\ell 0}$ along the $z$ axis of the box. This is sufficient, since only this projection contributes to the signal we discuss; see Fig. 1. We see distant clusters in their past, i.e., when the Universe was younger and the ISW effect was weaker. To take this into account we multiply the correlation function by the growth factor of the ISW effect. We obtain the correlation function for the SW effect in a similar manner, using a thin spherical layer as a convolution kernel. We do not take into account the variations of the last scattering surface radius with time, since we are interested in small distances, $<2000$~Mpc/h (the present radius of the last scattering surface is 9500~Mpc/h in the comoving coordinate frame). The obtained correlation functions of the ISW and SW effects are shown in
Fig. 4.

\begin{figure}[tbh]
  \includegraphics[width=1\columnwidth]{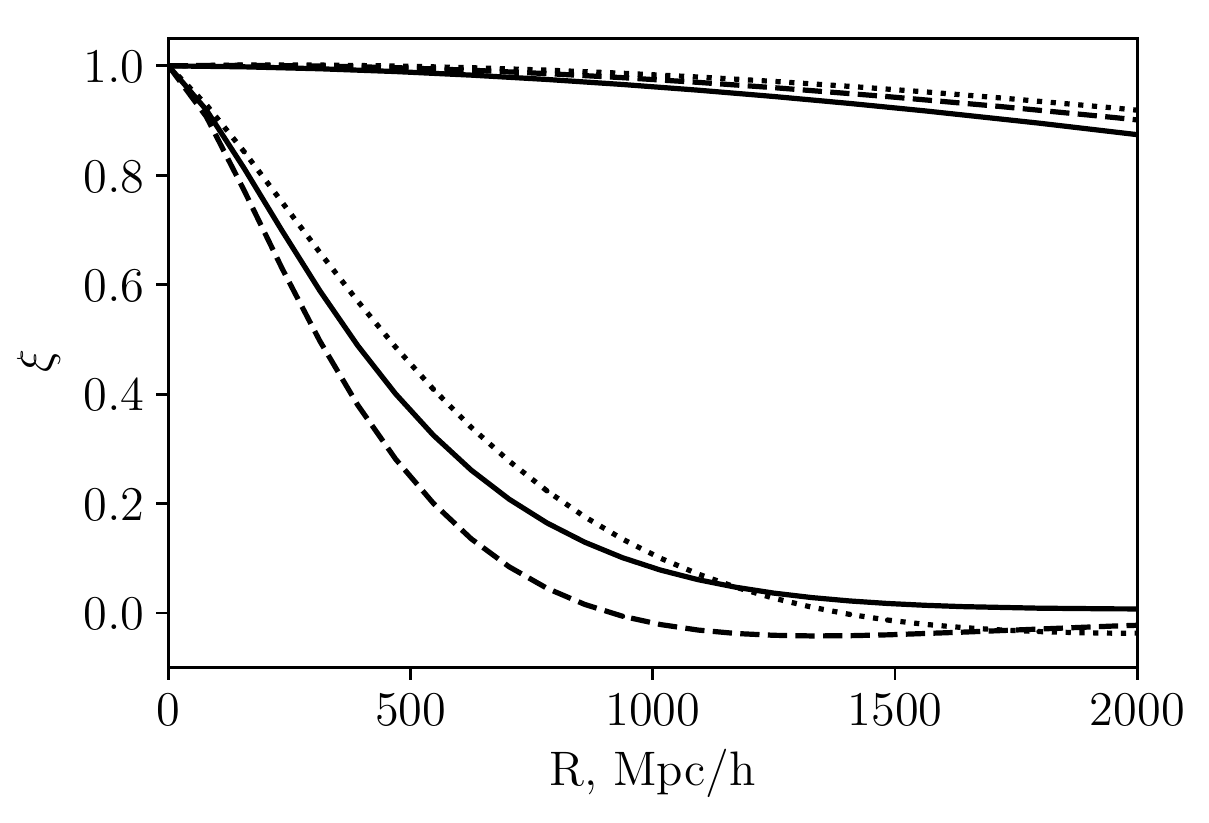}
  \caption{Correlation functions for dipole (dots), quadrupole (solid) and octupole (dashes) for the SW effect on the last scattering surface (top 3 lines) and the ISW effect (bottom three lines).}
\end{figure}

\begin{figure}[tbh]
  \includegraphics[width=1\columnwidth]{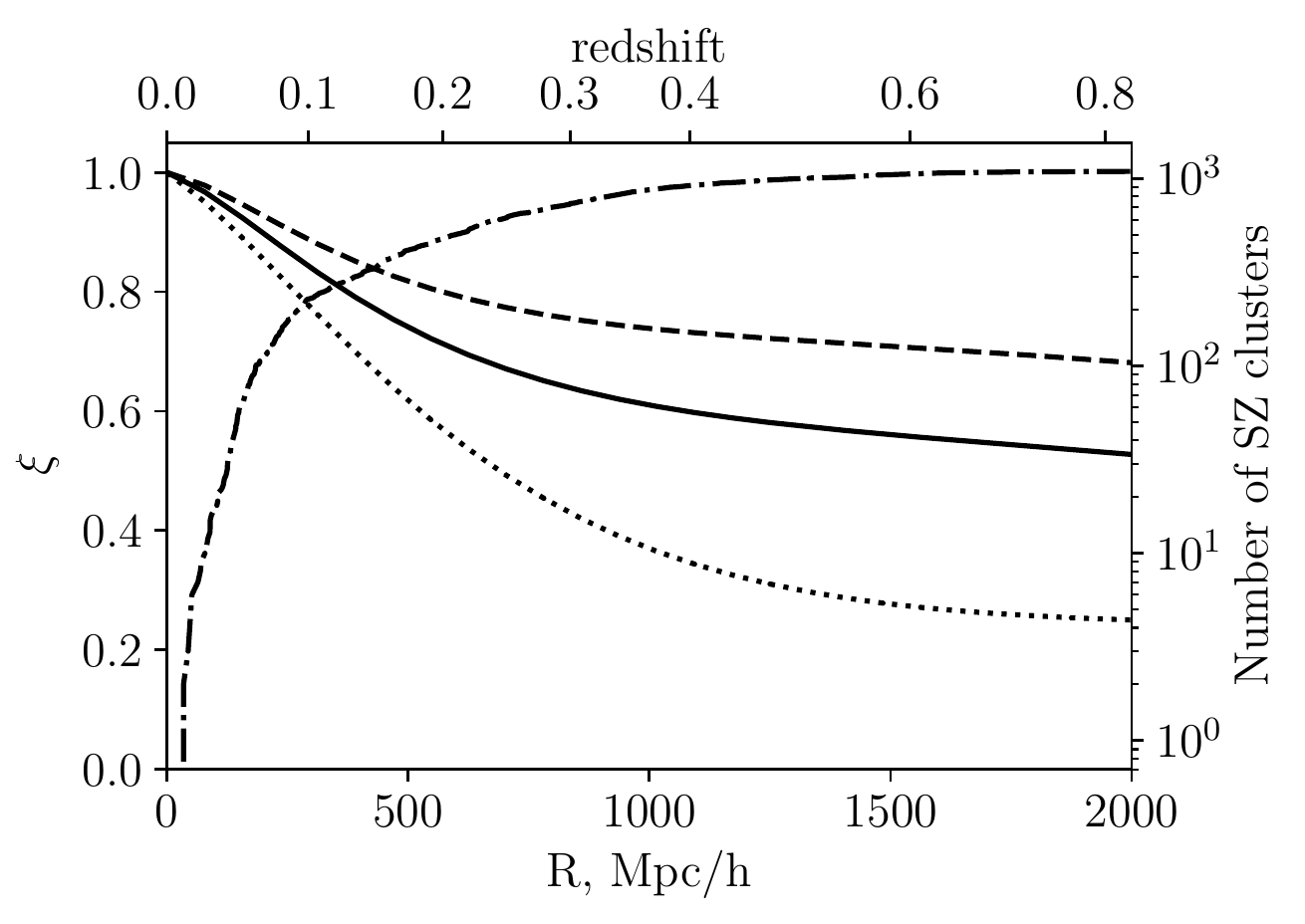}
  \caption{Total (SW+ISW) correlation functions for dipole (dots), quadrupole (solid) and octupole (dashes). The dot-dashed line shows the number of clusters at distance $<R$ from PLANCKSZ2 catalog \citep{2016A&A...594A..27P}.}
\end{figure}
    
    As one can see from Fig. 4, the correlation length for the SW effect is much larger than that for the ISW effect. This means that we have three regimes. For the clusters at distances smaller than the ISW correlation length we have $a_{\ell m}$ close to their local values. At distances much larger than the ISW correlation length we will detect $a_{\ell m}$ which have no ISW contribution, since we see a much younger Universe. Most SZ clusters are located closer than the SW correlation length, so they will share the same $a_{\ell m}$'s from the SW effect. At intermediate distances, comparable to the correlation length of the ISW effect, the amplitudes of spherical harmonics will fluctuate. The total correlation function of the $a_{\ell m}$'s detected by our method is shown in Fig. 5 together with cluster counts. One should note that it describes only the correlation of the local
    $a_{\ell m}$ with distant ones, not the correlation between two distant points.

 One can conclude from this analysis, that we have a unique opportunity to
 separate ISW and SW effects. Signals from distant clusters contain information about the
 SW effect only. Therefore dipole, quadrupole and octupole reconstructed by using only
 distant clusters will contain information about the CMB anisotropy directly at
 the surface of last scattering. In order to reconstruct total $a_{\ell m}$, $l=1,2,3$
 (SW+ISW effect) in our location we shall use only nearby clusters.    

\subsection{Precision of $a_{\ell m}$ $\ell=1,2,3$ reconstruction}

We estimate the relative precision for the reconstruction of different $a_{\ell m}$'s using measurements of CMB spectral distortions in the direction of SZ clusters. For simplicity we assume that $a_{\ell m}$'s are equal at the positions of clusters. This assumption should be valid in two cases, as discussed before: at very small distances, $R<250$~Mpc/h where we can measure the sum of the SW and ISW effects (i.e. $a_{\ell m}$'s are the same as measured directly from the CMB), or at large distances, $R>1000$~Mpc/h, where anisotropy is generated only by the SW effect. Assuming also equal and independent measurement errors of our signal given in Eq. (10), we obtain estimates of the errors of $a_{\ell m}$ measurements using the maximum likelihood method.

In Figs. 6 and 7 we show relative errors of 15 coefficients for 173 clusters at $R<250$~Mpc/h and 225 clusters at $R>1000$~Mpc/h. They are normalized by the largest error set to one. We use $4\pi$ normalized real spherical harmonics for this computation. One can see that amplitudes with $m=\pm \ell$ have the largest errors for every $\ell$. This happens because $Y_m^{\pm m}\sim sin^m(\theta)$.
Most of the power of these harmonics are close to the Galactic plane ($\theta\approx\pi/2$). This is the zone
of avoidance of Planck SZ catalog, where there are no clusters. In Fig. 8 we show how the relative errors for dipole, quadrupole and octupole powers ($C_1, C_2, C_3$) depend on the number of clusters or distance for the sample of $R>1000$ clusters.

From this analysis one can conclude, that to reliably recover all 15 components
of the three lowest multipoles the sensitivity should be at least one order of magnitude higher, than
that needed for the detection of the total signal.

\begin{figure}[tbh]
  \includegraphics[width=1\columnwidth]{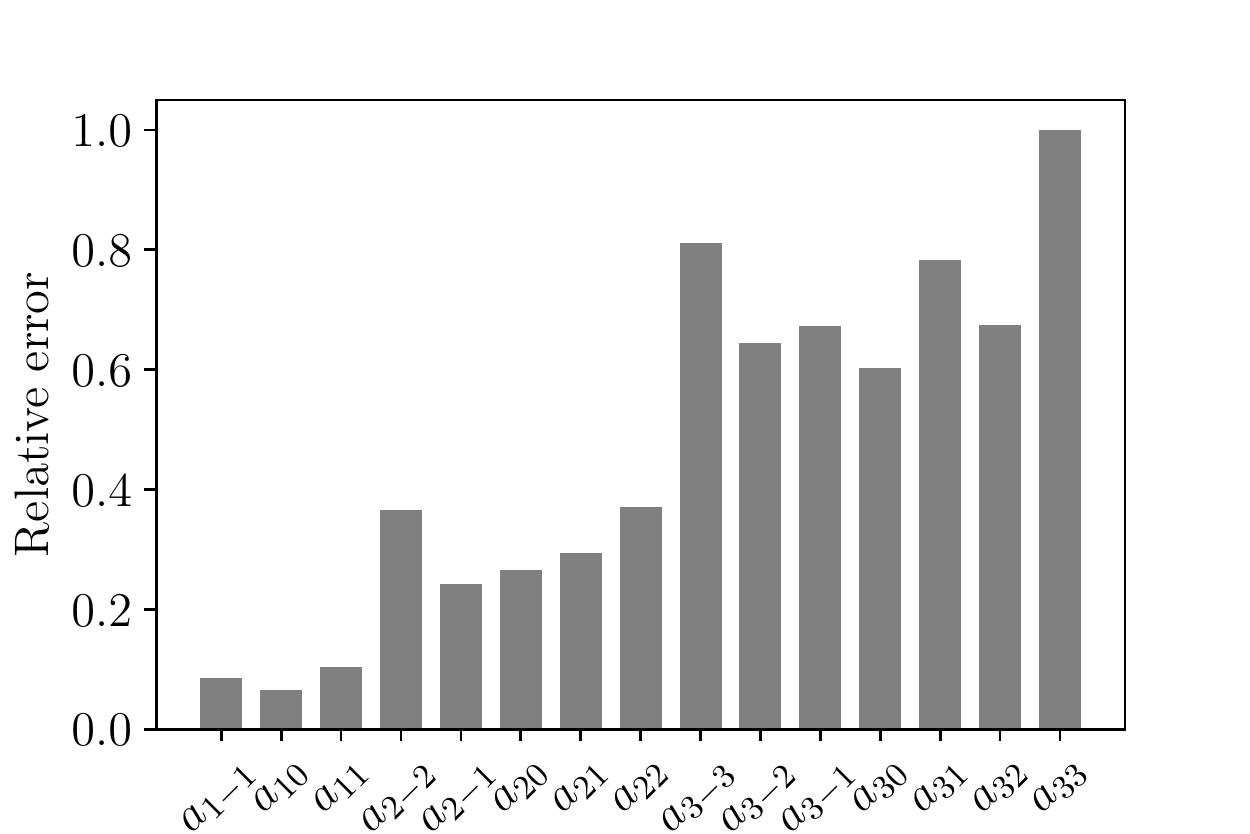}
  \caption{Relative errors of 15 multipole amplitudes measured from the CMB spectral distortions in SZ clusters at $R<250$~Mpc/h.}
\end{figure}

\begin{figure}[tbh]
  \includegraphics[width=1\columnwidth]{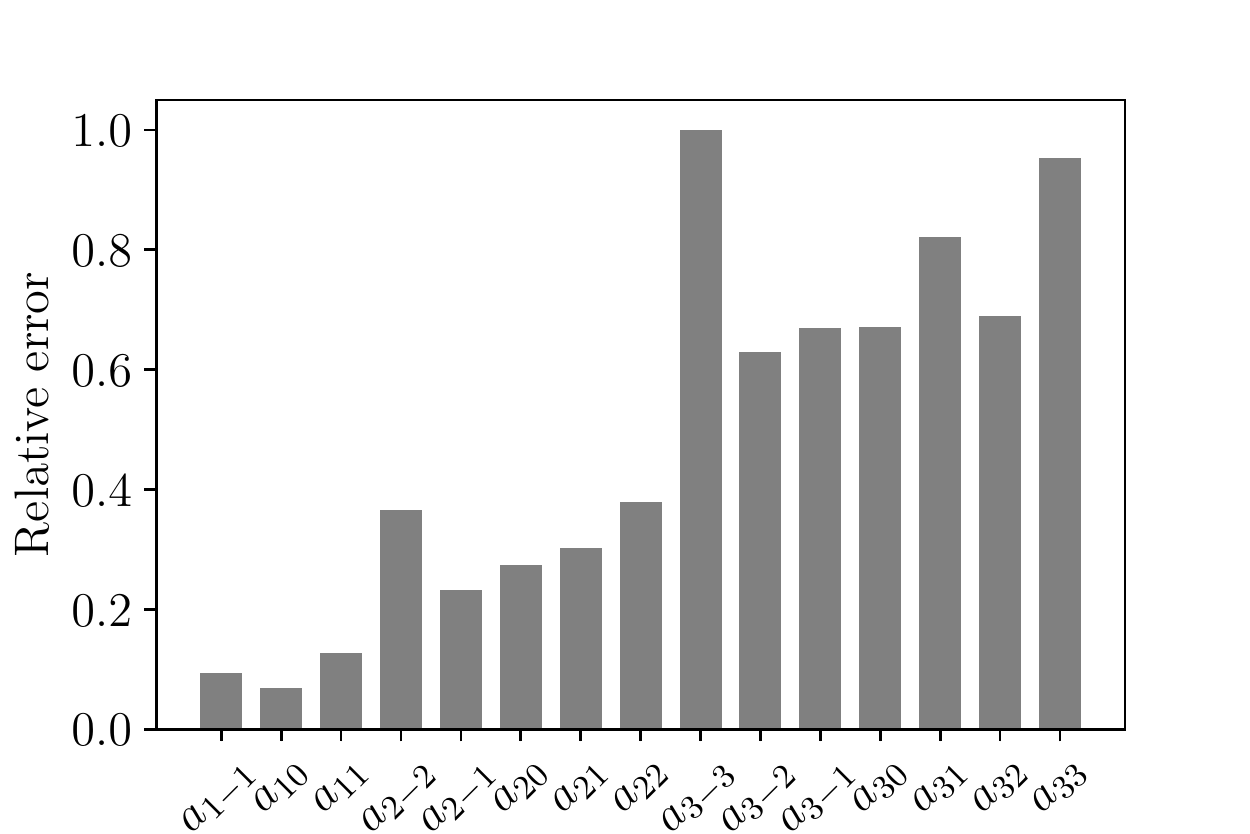}
  \caption{Relative errors of 15 multipole amplitudes measured from the CMB spectral distortions in SZ clusters at $R>1000$~Mpc/h.}
\end{figure}

\begin{figure}[tbh]
  \includegraphics[width=1\columnwidth]{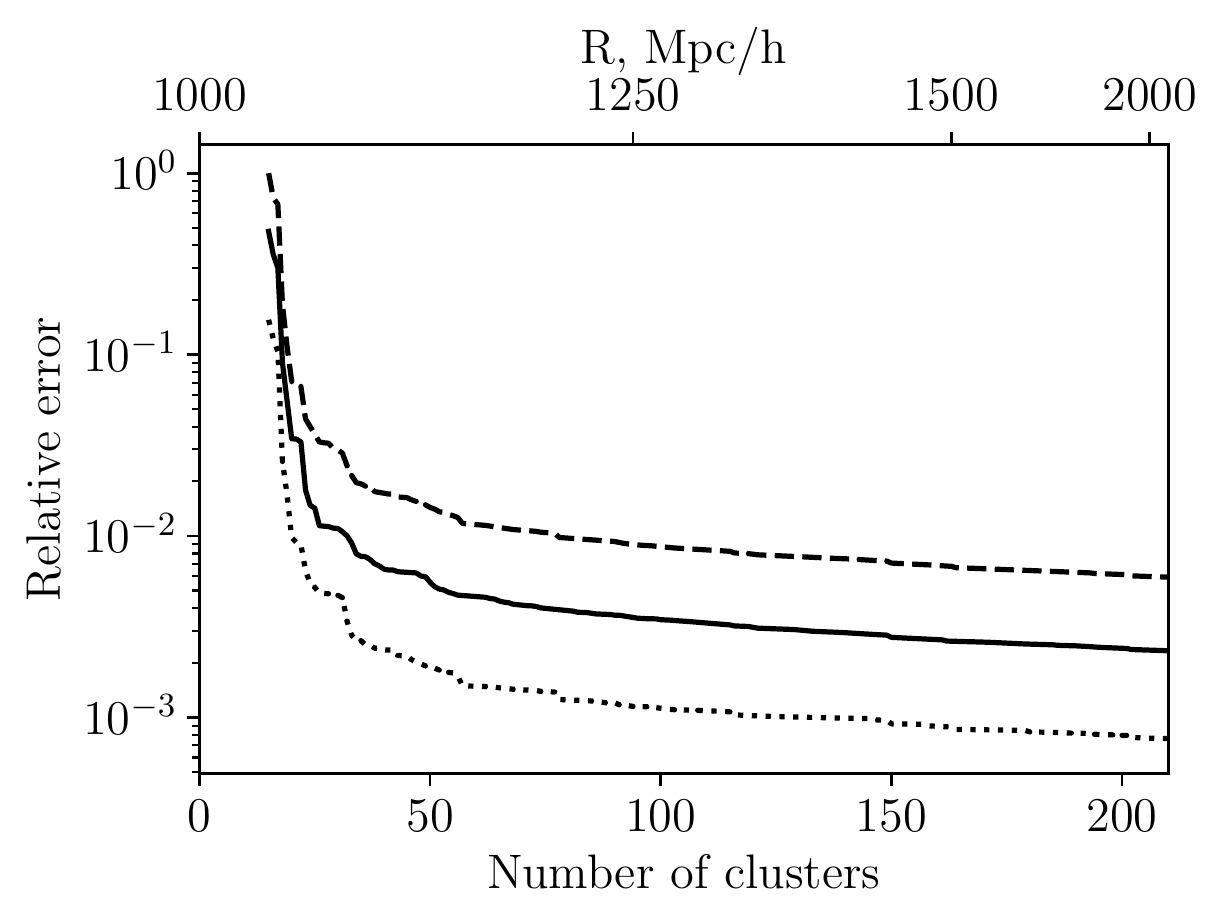}
  \caption{Relative error of the power of dipole (dots), quadrupole (solid) and octupole (dashed) for a sample of distant clusters as a function of distance or number of clusters (starting from 15 clusters).}
\end{figure}

\section{Conclusions}
In our paper we derived the correction to the thermal Sunyaev-Zel'dovich effect
due to the presence of $\ell=1,2,3$ anisotropy in CMB radiation.
The signal corresponding to
this correction has a very distinguishable features and can be separated from
other nonblackbody components. This signal is strong enough to be detected
by upcoming Millimetron mission. We have not considered in
details a possible way of components separation in the signal we about to
observe to identify the effect, that we described. We only have to mention, that in order to isolate the effect considered by us, it is necessary to take into account relativistic corrections to the isotropic thermal SZ effect at least up to the fifth order, the kinematic SZ effect and corrections
to it, as well as multiple scattering. In other words we should consider
all effects of the same and higher orders.
This correction makes it possible to independently estimate low the CMB  anisotropy multipoles by observing the spectral distortions from galaxy clusters.
We can use both nearby and distant clusters. The nearby clusters can be used
for independent estimations of all three local multipoles---dipole, quadrupole
and octupole---in our location. As for the distant clusters, they can be used to measure
the signal directly from the surface of last scattering without the
ISW contribution. Therefore the anisotropic thermal SZ effect gives us an opportunity
to separate the ISW and SW effects. 

We wish to thank the Referee for very helpful discussion.
The work is supported by the Program 28 of the fundamental research of the Presidium of the Russian Academy of Sciences ``Space: research of the fundamental processes and relations'', subprogram II ``Astrophysical objects as space laboratories'' and by the Project 01-2018 of LPI new scientific
groups.

\def\apj{Astrophys.~J}
\def\apjl{Astrophys.~J.,~Lett}
\def\apjs{Astrophys.~J.,~Supplement}
\def\an{Astron.~Nachr}
\def\aap{Astron.~Astrophys}
\def\mnras{Mon.~Not.~R.~Astron.~Soc}
\def\pasp{Publ.~Astron.~Soc.~Pac}
\def\aaps{Astron.~and Astrophys.,~Suppl.~Ser}
\def\apss{Astrophys.~Space.~Sci}
\def\ibvs{Inf.~Bull.~Variable~Stars}
\def\japa{J.~Astrophys.~Astron}
\def\na{New~Astron}
\def\aspproc{Proc.~ASP~conf.~ser.}
\def\aspcs{ASP~Conf.~Ser}
\def\aj{Astron.~J}
\def\actaa{Acta Astron}
\def\araa{Ann.~Rev.~Astron.~Astrophys}
\def\caosp{Contrib.~Astron.~Obs.~Skalnat{\'e}~Pleso}
\def\pasj{Publ.~Astron.~Soc.~Jpn}
\def\memsai{Mem.~Soc.~Astron.~Ital}
\def\astl{Astron.~Letters}
\def\aipproc{Proc.~AIP~conf.~ser.}
\def\physrep{Physics Reports}
\def\jcap{Journal of Cosmology and Astroparticle Physics}

\bibliography{asz.bib}



\end{document}